\documentclass[aps,prc,showpacs,nofootinbib,twocolumn]{revtex4-1}

\usepackage{epsfig}
\usepackage{graphicx,amsmath}
\usepackage{color}
\newcommand\ba{\begin{eqnarray}}
\newcommand\ea{\end{eqnarray}}

\newcommand{\be}{\begin{equation}}
\newcommand{\ee}{\end{equation}}
\newcommand{\bas}{\begin{eqnarray*}}
\newcommand{\eas}{\end{eqnarray*}}
\begin{document}
\title{\bf \large Use of nuclei fractal properties for help to determine some unknown particle spins and unknown nuclei excited level spins}

\author{B. Tatischeff$^{1,2}$\\
$^{1}$Univ Paris-Sud, IPNO, UMR-8608, Orsay, F-91405\\
$^{2}$CNRS/IN2P3, Orsay, F-91405}
\thanks{tati@ipno.in2p3.fr}
 
\vspace*{1cm}
\begin{abstract}
The fractal property stipulates that the same physical laws apply for different scales of a given physics. This property is applied to particles and nuclei, in order to study the possibility to use it to help for determination of unknown spins of some particles or excited nuclei levels.  
\end{abstract}
\maketitle
\section{Introduction}
 The fractal concept stipulates that the same physical laws apply for different scales of a given physics \cite{mandelbrot} \cite{nottale1}. The possibility to study several particle and nuclei properties with help of fracatals was recently shown. First, fractal properties were applied successively to fundamental force coupling constants, to atomic energies, and to elementary particle masses \cite{frac2}. Then, they were applied to study hadron spectroscopy \cite{frachadron}. Finally, they were applied to the study of masses and energy levels of several nuclei \cite{fracnoyau}.

The aim of the present paper, is to look at the possibility of using the same method to help to determine the unknown spins of some particles or nuclei excited levels.

The fractal concept, as well as the log-periodic corrections \cite{sornette} \cite{nottale} were summarized in \cite{fracnoyau}. The scale invariance is defined in the following way: an observable O(x), depending on the variable x,  is scale invariant under the arbitrary change x $\to~\lambda$x, if there is a number $\mu(\lambda)$ such that O(x) = $\mu$O($\lambda$x). $\lambda$ is the fundamental scaling ratio. The solution of  O(x) is the power law:
\be
 O(x) = C  x ^ { \alpha}
\ee
 where $\alpha$ = -ln$\mu$/ln$\lambda$.

We have therefore " a continuous translational invariance expressed on the logarithms of the variables" since 
\be
log(O(x)) = \alpha~log(x) + log(C)
\ee
 Considering a sequence of same objects, with a given spin,  O(x) is the mass of this object of rank "x". If the logs of the masses of a sequence having a given spin, are aligned  
versus the logs of the rank, we can look if the unknown spin mass can be introduced without loss of linearity in the straight line.  

It was shown \cite{frachadron} \cite{fracnoyau} that the masses follow discrete scale invariances and not continuous scale invariances; we observe therefore multi-fractal distributions.

This fractal property of linearity between the log of masses of a given spin versus the log of the rank is used below to help to determine the unknown spins of some mesons, hadrons and nuclei levels. 
When an unknown spin level arrives at the same rank for two (or more) different spin distributions, only the larger spin appears in the figures, the other lower spin marks being recovered. 
\section{Mesons}
The masses and spins are the last Particle Data Group values (PDG) \cite{pdg}. Some masses, omitted from the summary table of \cite{pdg}, are however kept in the study. Some meson families are ignored when a too small number of masses or spins is known.
\subsection{Unflavoured mesons}
\begin{figure}[ht]
\begin{center}
\hspace*{-3.mm}
\vspace*{5.mm}
\scalebox{1.05}[1]{
\includegraphics[bb=46 242 526 550,clip,scale=0.5]{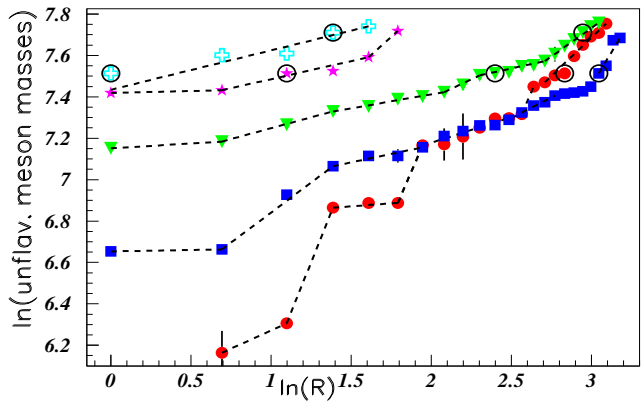}}
\caption{Color on line. Log-log plot of unflavoured meson  masses. Full red circles correspond to J = 0 mesons,  full blue squares correspond to J = 1 mesons, full green triangles correspond to J = 2 meson,  full purple stars
correspond to J = 3 mesons, and sky blue empty crosses correspond to J = 4 mesons (see text).}
\end{center}
\end{figure}
The log-log plot of unflavoured meson  masses is shown in figure~1. 
Full red circles correspond to J = 0 mesons,  full blue squares correspond to J = 1 mesons, full green triangles correspond to J = 2 mesons,  full purple stars
correspond to J = 3 mesons, and sky blue empty crosses correspond to J = 4 mesons. The first (pion) mass is omitted in order to expand the figure scale. The mass and width of the $f_{0}$(600) are reduced since the first 14 entries in PDG table 2011, give its mass to be in the range 440-552~MeV. Therefore, I kept the following value: M = 475$\pm$50~MeV. All
unflavoured mesons with spin = 0, 1, 2, 3, or 4, therefore up to $f_{2}$(2340), are kept in figure~1.  The large number of observed unflavoured mesons, as well as their known spins, allows to discuss the possibilities and the limitations of the method.
  
The different straight lines indicate the multi-fractal properties. The first unknown spin meson is the X(1835) which mass mark (M = 1833.7~MeV, 
ln(M) = 7.514) is encircled and is introduced for all possible spins. The alignements must be looked for, ignoring these encircled marks. We conclude therefore, using these alignements, that J = 4 and J = 0, are unlikely for this meson. It is not possible to choose between spins J = 1, 2, or 3. The spin of the M = 2231.1~MeV unflavoured meson (ln(M) = 7.710) is noted to be either 2 or 4. The corresponding marks are again encircled.
The figure shows again that it is not possible to use this method to choose between both possible spins. 
\subsection{Strange mesons}
The log-log plot of strange meson masses is shown in figure~2. Full red circles correspond to J = 0 mesons,  full blue squares correspond to J = 1 mesons,  full green triangles correspond to J = 2 mesons, full purple stars correspond to J = 3 mesons, and sky blue empty crosses correspond to J = 4 mesons. Only one J~=~5 strange meson is known in the mass range studied. The spins of two meson masses M~=~1630~MeV, ln(M) = 7.396, and 3100~MeV, ln(M) = 8.039, are unknown. In the figure, these masses are introduced in all spin distributions and are encircled. From figure~2 we deduce that the spin of the strange meson with mass M = 1630~MeV, is likely to be J = 0. J = 2 or J = 4 are also possible, but not  J = 1 neither J = 3. Indeed, after removing the M~=~1630~MeV strange meson from the J~=~1 distribution, the two last J = 1 marks have to be translated to left by one step, and then the alignement is no more obtained for larger masses. The attribution of the spin to the strange meson at
M~=~3100~MeV is difficult, since it is not possible to exclude that new strange meson(s) could be observed in the range 2500$\le$M$\le$3100~MeV. From the present data J~=~3 is the most probable spin value.
\begin{figure}[ht]
\begin{center}
\hspace*{-3.mm}
\vspace*{5.mm}
\scalebox{1.05}[1]{
\includegraphics[bb=45 235 530 550,clip,scale=0.5]{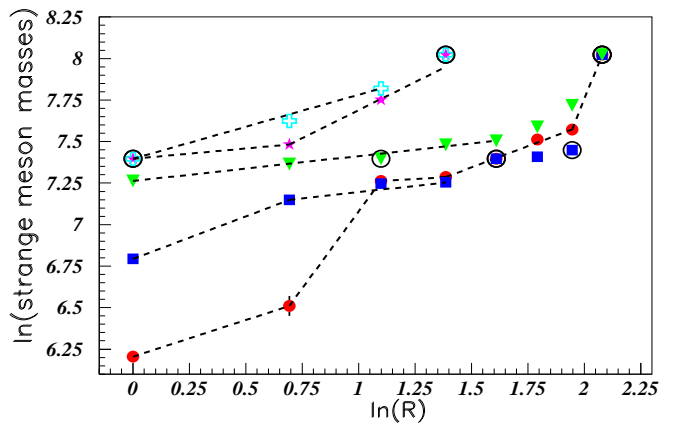}}
\caption{Color on line. Log-log plot of strange meson  masses. Full red circles correspond to J = 0 mesons,  full blue squares correspond to J = 1 mesons, full green triangles correspond to J = 2 mesons, full purple stars correspond to J = 3 mesons, and sky blue empty crosses correspond to J = 4 mesons (see text).}
\end{center}
\end{figure}
When the mass unprecision is not given, a value of $\Delta$M~=~50~MeV is used. 
\subsection{Charmed mesons}
\begin{figure}[ht]
\begin{center}
\hspace*{-3.mm}
\vspace*{5.mm}
\scalebox{1.05}[1]{
\includegraphics[bb=45 235 530 550,clip,scale=0.5]{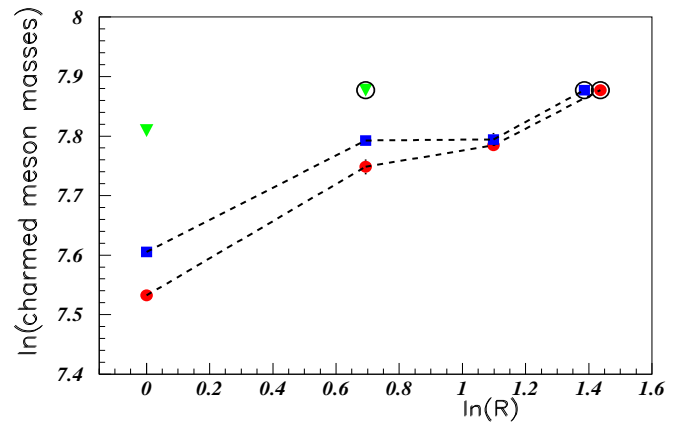}}
\caption{Color on line. Log-log plot of charmed meson  masses. Full red circles correspond to J = 0 mesons,  full blue squares correspond to J = 1 mesons, and full green triangles correspond to J = 2 mesons (see text).}
\end{center}
\end{figure}
The log-log plot of charmed  meson masses is shown in figure~3. Full red circles correspond to J = 0 mesons,  full blue squares correspond to J = 1 mesons, and full green triangles correspond to J = 2 mesons. The first charmed meson with unknown spin is the D$^{*}$ at M = 2637, ln(M) = 7.877. The corresponding marks are introduced in all possible spin data, and  are encircled. Here again, the method discussed is unable to help us to conclude firmly, although J = 0 is weakly favoured.
\subsection{Charmed, strange mesons}
\begin{figure}[ht]
\begin{center}
\hspace*{-3.mm}
\vspace*{5.mm}
\scalebox{1.05}[1]{
\includegraphics[bb=43 235 530 550,clip,scale=0.5]{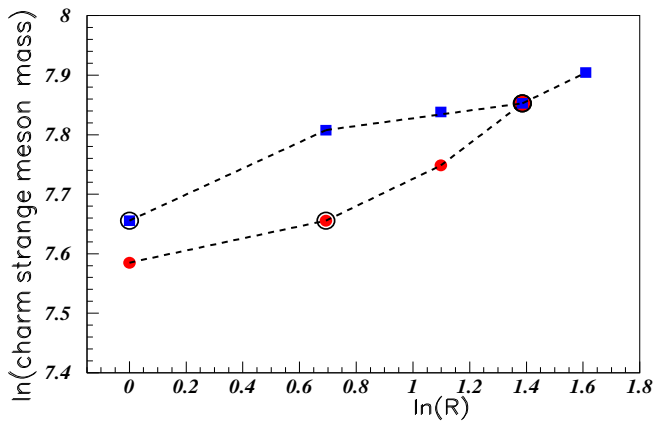}}
\caption{Color on line. Log-log plot of charmed strange meson masses. Full red circles correspond to J = 0 mesons and full blue squares correspond to J = 1 mesons (see text).}
\end{center}
\end{figure}
The log-log plot of charmed, strange  meson masses is shown in figure~4. Full red circles correspond to J = 0 mesons and full blue squares correspond to J = 1 mesons. The spin of the second charmed, strange meson D$_s$(2112.3~MeV) (log(M)~=~7.6555) is unknown. The figure shows that it is not possible to suggest a spin value for it. The spin of the 6$^{th}$ charmed, strange meson D$_{s2}$(2572.6~MeV) (log(M)~=~7.8527) is also unknown. Figure~4 shows that it is possible to suggest J = 1 for its spin.
\subsection{$c - {\bar c}$ mesons}
Many $c - {\bar c}$ meson masses are observed but, above M = 3900~MeV, nearly all spins are undetermined. Therefore we will only consider here the study of log-log distributions up to M = 3930~MeV. Figure~5 shows the log-log plot of $c - {\bar c}$  meson masses. Full red circles correspond to J = 0 mesons,  full blue squares correspond to J = 1 mesons, and full green triangles correspond to J = 2 mesons. The spins of the M = 3871.56, ln(M) = 8.261, and 3915.5~MeV, ln(M) = 8.273 $c - {\bar c}$ mesons are not known. These masses are therefore introduced in the three distributions corresponding to J~=~0, 1, or 2 and are encircled. From the figure, we get two posssible spins for the M = 3871.56 
$c - {\bar c}$ meson, namely J = 0 or 1, and also preferably one spin: J=1 for the M = 3915.5 
$c - {\bar c}$ meson. It is however not possible to eliminate J = 2 for both levels, but the small number of $c - {\bar c}$  mesons with such spin makes such conclusion doubtful.
\begin{figure}[ht]
\begin{center}
\hspace*{-3.mm}
\vspace*{5.mm}
\scalebox{1.05}[1]{
\includegraphics[bb=43 240 520 550,clip,scale=0.5]{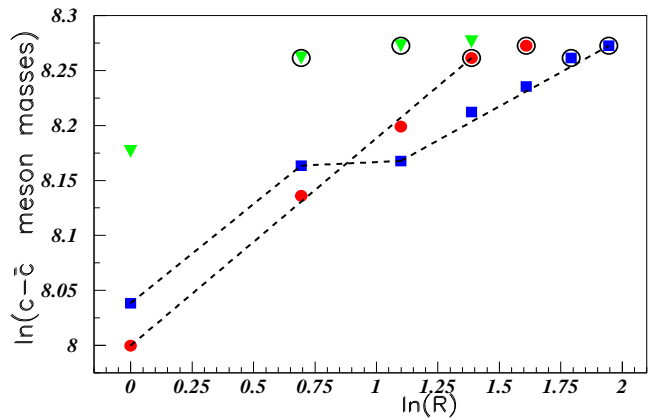}}
\caption{Color on line. Log-log plot of $c - {\bar c}$  masses. Full red circles correspond to J = 0 mesons,  full blue squares correspond to J = 1 mesons, and full green triangles correspond to J = 2 mesons (see text).}
\end{center}
\end{figure}
\section{Baryons}
\subsection{N baryons}
\begin{figure}[ht]
\begin{center}
\hspace*{-3.mm}
\vspace*{5.mm}
\scalebox{1.05}[1]{
\includegraphics[bb=55 235 520 550,clip,scale=0.5]{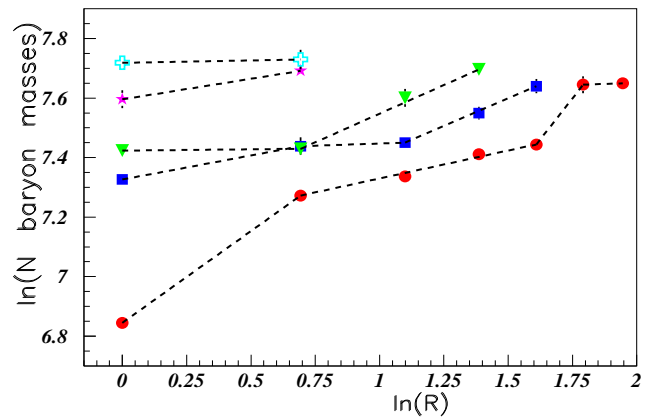}}
\caption{Color on line. Log-log plot of N baryon  masses. Full red circles correspond to J = 1/2 baryons,  full blue squares correspond to J = 3/2 baryons, full green triangles correspond to J = 5/2 baryons, full purple stars
correspond to J = 7/2 baryons, and sky blue empty crosses correspond to J = 9/2 (see text).}
\end{center}
\end{figure}
Figure~6 shows the log-log plot of N baryons. Full red circles correspond to J = 1/2 baryons,  full blue squares correspond to J = 3/2 baryons, full green triangles correspond to J = 5/2 baryons, full purple stars
correspond to J = 7/2 baryons, and sky blue empty crosses correspond to J = 9/2. All such baryons have known spin. The figure shows many straight lines,  a signature of multi-fractal properties. 
\subsection{$\Delta$ baryons}
\begin{figure}[ht]
\begin{center}
\hspace*{-3.mm}
\vspace*{5.mm}
\scalebox{1.05}[1]{
\includegraphics[bb=55 235 520 550,clip,scale=0.5]{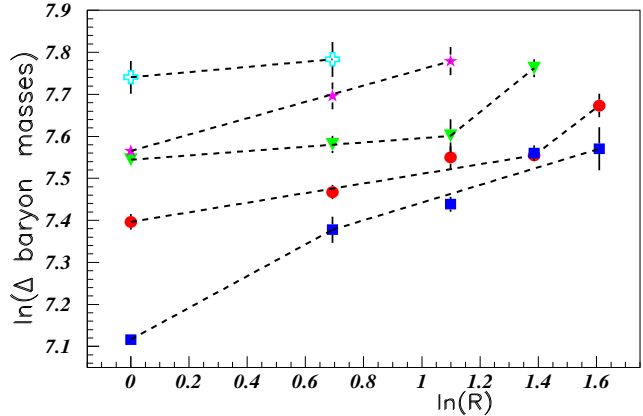}}
\caption{Color on line. Log-log plot of $\Delta$ baryon  masses. Full red circles correspond to J = 1/2 baryons,  full blue squares correspond to J = 3/2 baryons, full green triangles correspond to J = 5/2 baryons, full purple stars
correspond to J = 7/2 baryons, and sky blue empty crosses correspond to J = 9/2 baryons (see text).}
\end{center}
\end{figure}
Figure~7 shows the log-log plot of $\Delta$ baryons. Full red circles correspond to J = 1/2 baryons,  full blue squares correspond to J = 3/2 baryons, full green triangles correspond to J = 5/2 baryons, purple stars
correspond to J = 7/2 baryons, and sky blue empty crosses correspond to J = 9/2 baryons. All such baryons have known spin. The figure shows - again - many straight lines,  a signature of multi-fractal properties. 
\subsection{$\Lambda$ baryons}
\begin{figure}[ht]
\begin{center}
\hspace*{-3.mm}
\vspace*{5.mm}
\scalebox{1.05}[1]{
\includegraphics[bb=52 235 520 550,clip,scale=0.5]{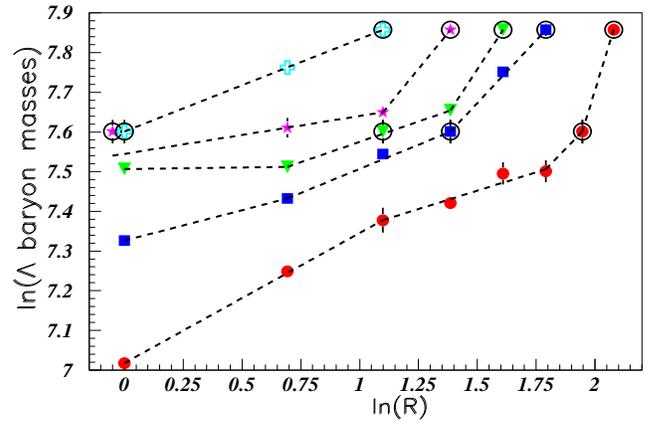}}
\caption{Color on line. Log-log plot of $\Lambda$ baryon  masses. Full red circles correspond to J = 1/2 baryons,  full blue squares correspond to J = 3/2 baryons, full green triangles correspond to J = 5/2 baryons, full purple stars
correspond to J = 7/2 baryons, and sky blue empty crosses correspond to J = 9/2 baryons (see text).}
\end{center}
\end{figure}
Figure~8 shows the log-log plot of $\Lambda$ baryons. Full red circles correspond to J = 1/2 baryons,  full blue squares correspond to J = 3/2 baryons, full green triangles correspond to J = 5/2 baryons, full purple stars
correspond to J = 7/2 baryons, and sky blue empty crosses correspond to J = 9/2. The spin of two $\Lambda$ baryons is unknown, namely the $\Lambda$(2000) (ln(M) = 7.60) and  $\Lambda$(2585) (ln(M) = 7.8575). The corresponding marks in figure~8 are encircled. From the figure, we conclude that the spin of $\Lambda$(2000) is unlikely to be J = 1/2 neither 7/2, but rather J = 3/2,  5/2, or 9/2. Two spin values are preferred for $\Lambda$(2585) J = 3/2 or 5/2.

Two spins are preferred for the $\Lambda$(2585), J = 3/2 or 9/2.
\subsection{$\Sigma$ baryons}
\begin{figure}[ht]
\begin{center}
\hspace*{-3.mm}
\vspace*{5.mm}
\scalebox{1.05}[1]{
\includegraphics[bb=46 236 516 545,clip,scale=0.5]{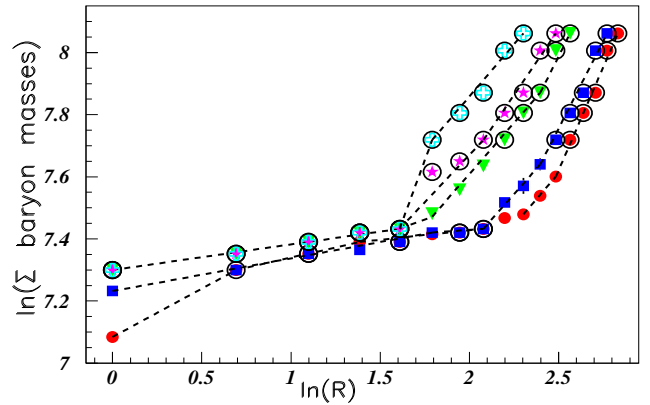}}
\caption{Color on line. Log-log plot of $\Sigma$ baryon  masses. Full red circles correspond to J = 1/2 baryons,  full blue squares correspond to J = 3/2 baryons, full green triangles correspond to J = 5/2 baryons, purple stars
correspond to J = 7/2 baryons, and sky blue empty crosses correspond to J = 9/2 baryons (see text).}
\end{center}
\end{figure}
Figure~9 shows the log-log plot of $\Sigma$ baryons. Full red circles correspond to J = 1/2 baryons,  full blue squares correspond to J = 3/2 baryons, full green triangles correspond to J = 5/2 baryons, full purple stars
correspond to J = 7/2 baryons, and sky blue empty crosses correspond to J = 9/2 baryons. The spins of many $\Sigma$ baryons are unknown. The corresponding marks are encircled. There is no known $\Sigma$ baryon with spin J = 9/2, and only one at M = 2030~MeV (ln(M) = 7.616), is known to have J = 7/2.
Since this last baryon mark is ouside the continuity of other unknown spin baryons, the unknown spin of nearby baryons should be different from J = 7/2. The too large number of  $\Sigma$ baryons with unknown spin, prevents the use of this method to tentatively predict possible spins.

The same conclusion applyed for $\Xi$, $\Omega$, and more generally for all heavier  baryons, like charmed or bottom.
\section{Nuclei}
In this section, the method is applied to a limitated selection of nuclei.
\subsection{The $^{4}$He nucleus}
The masses and spins of the energy levels of $^{4}$He are known up to E$_{x}$ = 28.5~MeV \cite{meyer}.
\begin{figure}[ht]
\begin{center}
\hspace*{-3.mm}
\vspace*{5.mm}
\scalebox{1.05}[1]{
\includegraphics[bb=46 237 530 550,clip,scale=0.5]{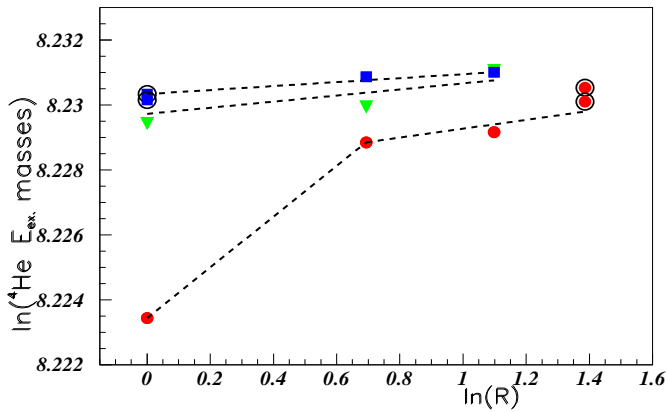}}
\caption{Color on line. Log-log plot of $^{4}$He E$_{ex.}$ masses (see text). Full red circles correspond to J = 0 levels,  full blue squares correspond to J = 1 levels, and full green triangles correspond to J = 2 levels (see text).}
\end{center}
\end{figure}
Figure~10 shows the log-log plot of the $^{4}$He E$_{ex.}$ masses \cite{meyer}. Full red circles correspond to J = 0 levels,  full blue squares correspond to J = 1 levels, and full green triangles correspond to J = 2 levels. In the range 24.9$\le~E_{ex.}~\le$26.5~MeV, there is a 
($J^{P},T$) ($1^{-}, 1$) level at a poorly defined mass, either at M = 25.8~MeV, or M = 25.1~MeV. In the same way, a ($0^{-}, 1$) level is reported either at M = 26.5~MeV, or at M = 24.9~MeV. The corresponding marks are encircled. Figure~10 shows that the lower mass is preferred for the ($0^{-}, 1$) level. Both masses for the ($1^{-}, 1$) level are too close to allow making a preference between them.
\subsection{The $^{9}$Be nucleus}
Figure~11 shows the log-log plot of $^{9}$Be masses \cite{tilley} versus the corresponding rank "R".
The masses and spins of the energy levels of $^{9}$Be are known up to E$_{x}$ = 13~MeV \cite{tilley}. 
\begin{figure}[ht]
\begin{center}
\hspace*{-3.mm}
\vspace*{5.mm}
\scalebox{1.}[1]{
\includegraphics[bb=19 236 520 550,clip,scale=0.5]{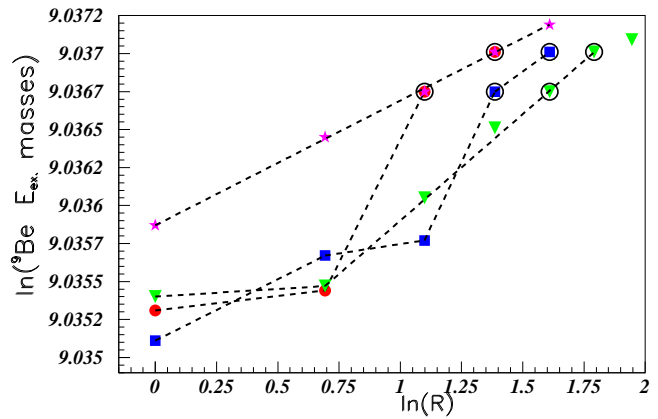}}
\caption{Color on line. Log-log plot of $^{9}$Be masses. Full red circles correspond to J = 1/2 levels,  full blue squares correspond to J = 3/2 levels, full green triangles correspond to J = 5/2 levels, and full purple stars correspond to J = 7/2 levels (see text).}
\end{center}
\end{figure}
The full red circles, full blue squares, full green triangles, and full purple stars correspond respectively to the spin values: J=1/2, 3/2, 5/2, and 7/2. The first unknown spin level, at E$_{x}$ = 13.79~MeV, ln(M) = 2.624, is tentatively introduced inside all four possible spins, by encircled points. The second unknown spin level, at  E$_{x}$ = 15.97~MeV, ln(M) = 2.771, is also tentatively introduced inside all four possible spins, by encircled points. 

We observe that the linearity is obtained for both unknown spin levels, for the spin value J = 7/2, and is also extended to the next   J = 7/2 level at E$_{x}$ = 17.493~MeV, ln(M) = 2.862. 
There is no linearity for the spin J = 1/2, neither 3/2. The situation for J = 5/2 is more complicated. Indeed, if its log-log distribution is aligned up to both unknown spin levels, this is no more observed with the next level at E$_{x}$ = 16.671~MeV, ln(M) = 2.814. We notice that if both unknown spin levels have different spins, the linearities are broken  for J = 7/2, as well as for J = 5/2. But we notice however, that if the first unknown spin level is not J = 3/2, the second can have such spin J = 3/2.

The result of the used method is to tentatively attribute J = 7/2 for both levels of $^{9}$Be at E$_{x}$ = 13.79~MeV and E$_{x}$ = 15.97~MeV.
\subsection{The $^{13}$C nucleus}
\begin{figure}[ht]
\begin{center}
\hspace*{-3.mm}
\vspace*{5.mm}
\scalebox{1.}[1]{
\includegraphics[bb=24 238 520 545,clip,scale=0.5]{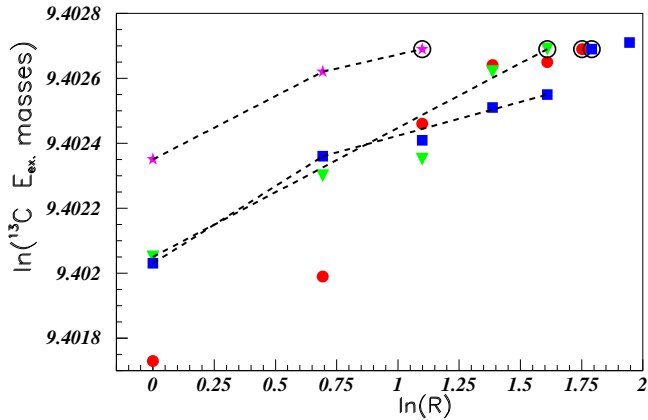}}
\caption{Color on line. Log-log plot of $^{13}$C masses. Full red circles correspond to J = 1/2 levels,  full blue squares correspond to J = 3/2 levels,  full green triangles correspond to J = 5/2 levels, and full purple stars correspond to J = 7/2 levels (see text).} 
\end{center}
\end{figure}
Figure~12 shows the log-log plot of $^{13}$C masses \cite{as1315} versus the corresponding rank "R". 
Full red circles correspond to J = 1/2 levels,  full blue squares correspond to J = 3/2 levels,  full green triangles correspond to J = 5/2 levels, and full purple stars correspond to J = 7/2 levels.
The level at M$_{ex.}$~=~11.748~MeV, ln(M) = 2.464, is the first one with unknown spin. It is tentatively introduced inside all four lower possible spins, by encircled points. None spin allows to draw a linear straight line between the log of the massses versus the log of the rank. Although it is therefore not possible to conclude, we notice that the (green) triangles, which correspond to J = 5/2, would exhibit a good linearity, if the third J = 5/2 mass would be somewhat larger.  On the other hand, the three last marks corresponding to J = 1/2 are aligned. 
\subsection{The $^{14}$N nucleus}
\begin{figure}[ht]
\begin{center}
\hspace*{-3.mm}
\vspace*{5.mm}
\scalebox{1.}[1]{
\includegraphics[bb=31 244 520 545,clip,scale=0.5]{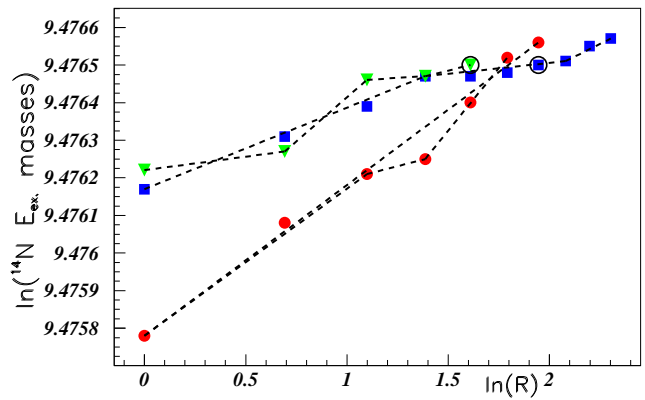}}
\caption{Color on line. Log-log plot of $^{14}$N masses (see text). Full red circles correspond to J = 1 levels,  full blue squares correspond to J = 2 levels, and full green triangles correspond to J = 3 levels.} 
\end{center}
\end{figure}
Figure~13 shows the log-log plot of $^{14}$N masses \cite{as1315} versus the corresponding rank "R". 
Full red circles correspond to J = 1 levels,  full blue squares correspond to J = 2 levels, and full green triangles correspond to J = 3 levels. 
The level at M$_{ex.}$~=~9.386~MeV, ln(M) = 2.239, is the first one with unknown spin since it is reported as 
J$^{\pi}$(2$^{-}$, 3$^{-}$). Its mark is encircled in figure~13.  In view of the figure, the spin J = 2 is preferred for the M$_{ex.}$~=~9.386~MeV level. 
\subsection{The $^{15}$N nucleus}
\begin{figure}[ht]
\begin{center}
\hspace*{-3.mm}
\vspace*{5.mm}
\scalebox{1.}[1]{
\includegraphics[bb=25 245 520 550,clip,scale=0.5]{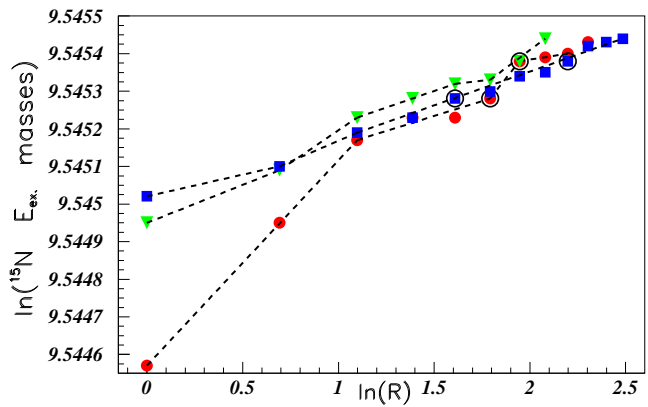}}
\caption{Color on line. Log-log plot of $^{15}$N masses (see text). Full red circles correspond to J = 1/2 levels,  full blue squares correspond to J = 3/2 levels, and full green triangles correspond to J = 5/2 levels. }
\end{center}
\end{figure}
Figure~14 shows the log-log plot of $^{15}$N masses \cite{as1315} versus the corresponding rank "R". 
Full red circles correspond to J = 1/2 levels,  full blue squares correspond to J = 3/2 levels, and full green triangles correspond to J = 5/2 levels. 
The level at M$_{ex.}$~=~9.928~MeV, ln(M) = 2.295, is the first one with unknown spin, it is reported as J$^{\pi}$(1/2, 3/2)$^{+}$. Another level at M$_{ex.}$~=~11.235~MeV, ln(M) = 2.419, has an unknown spin, since it is noted as 
J $\ge$ 3/2. The marks corresponding to the possible spins are encircled in figure~14. It is not easy to conclude firmly on the preferred spins for both levels. Indeed all four encircled marks lie on straight lines. We notice however the long straight line, including both unknown spin levels of the $^{15}$N nuclei for J = 3/2. This spin value is therefore preferred for the two M$_{ex.}$~=~9.928~MeV and 
M$_{ex.}$~=~11.235~MeV levels.
\subsection{The $^{16}$N nucleus}
\begin{figure}[ht]
\begin{center}
\hspace*{-3.mm}
\vspace*{5.mm}
\scalebox{1.}[1]{
\includegraphics[bb=22 236 515 550,clip,scale=0.5]{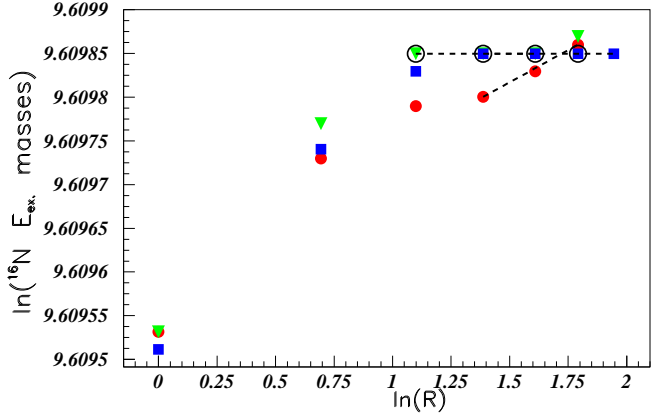}}
\caption{Color on line. Log-log plot of $^{16}$N masses (see text). Full red circles correspond to J = 1 levels,  full blue squares correspond to J = 2 levels, and full green triangles correspond to J = 3 levels.}
\end{center}
\end{figure}
Figure~15 shows the log-log plot of $^{16}$N masses \cite{as1617} versus the corresponding rank "R". 
Full red circles correspond to J = 1 levels,  full  blue squares correspond to J = 2 levels, and full green triangles correspond to J = 3 levels. A single J = 0 level at 
E$_{ex}$ = 0.1204~MeV is reported in the studied mass range. 

The spin of the level at 
E$_{ex}$ = 5.129~MeV, ln(M) = 1.635, is reported as J$\ge $ 2, and the spin of the level at 
E$_{ex}$ = 5.150~MeV, ln(M) = 1.639, is reported as J$^{P}$ = (2, 3)$^{-}$. Both level marks  are introduced in the figure~15. They are the third and fourth marks for the J = 3, and the fifth and sixth marks for J = 2.  We observe the straight line in this mass range since the increases of the excited state level masses are very small. The presence of the next 
J = 2 level mass at  E$_{ex}$ = 5.25~MeV, ln(M) = 1.658, brings us to privilege J = 2 for the  
E$_{ex}$~=~5.150~MeV level. Both values J = 2 and 3 are likely for the 
E$_{ex}$~=~5.129~MeV level.
\subsection{The $^{17}$F  nucleus}
\begin{figure}[ht]
\begin{center}
\hspace*{-3.mm}
\vspace*{5.mm}
\scalebox{1.}[1]{
\includegraphics[bb=30 242 520 550,clip,scale=0.5]{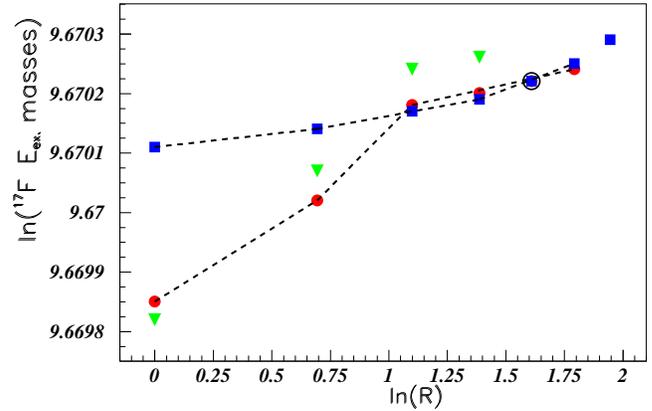}}
\caption{Color on line. Log-log plot of $^{17}$F excited level masses (see text). Full red circles correspond to J = 1/2 levels,  full blue squares correspond to J = 3/2 levels, and full green triangles correspond to J = 5/2 levels.}
\end{center}
\end{figure}
Figure~16 shows the log-log plot of $^{17}$F masses \cite{as1617} versus the corresponding rank "R". Full red circles correspond to J = 1/2 levels,  full blue squares correspond to J = 3/2 levels, and full green triangles correspond to J = 5/2 levels.
 The spin of the level at 
E$_{ex}$ = 6.406~MeV, ln(M) = 1.857, is reported as J$^{P}$ = (1/2, 3/2)$^{-}$. This mark of this level is encircled in figure~16. Here again, this method cannot be used to help to choose between both spin values.
\subsection{The $^{20}$F nucleus}
Figure~17 shows the log-log plot of $^{20}$F  excited level masses \cite{as1820} versus the corresponding rank "R". 
\begin{figure}[ht]
\begin{center}
\hspace*{-3.mm}
\vspace*{5.mm}
\scalebox{1.}[1]{
\includegraphics[bb=25 128 520 550,clip,scale=0.5]{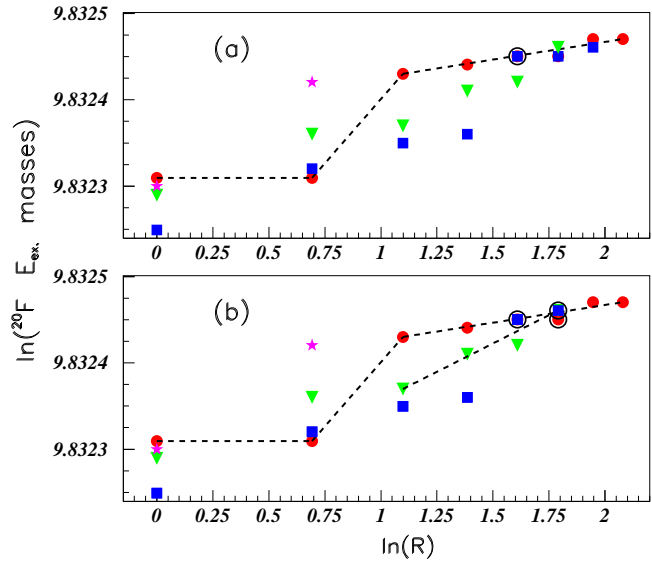}}
\caption{Color on line. Log-log plot of $^{20}$F excited level masses (see text). Full red circles correspond to J = 1 levels,  full blue squares correspond to J = 2 levels,  full green triangles squares correspond to J = 3 levels, and full pink stars correspond to J = 4 levels (see text).}
\end{center}
\end{figure}
 Full red circles correspond to J = 1 levels,  full blue squares correspond to J = 2 levels,  full green triangles squares correspond to J = 3 levels, and full pink stars correspond to J = 4 levels. In the range studied there is only one J = 0 level at E$_{ex}$ = 3.526~MeV, and one J = 5 level at  E$_{ex}$ = 1.8244~MeV. 
In the range studied here, there are three levels with uncertain spin, at:  E$_{ex}$ = 3.587~MeV, ln(M) = 1.277, E$_{ex}$ = 3.680~MeV, ln(M) = 1.303, and E$_{ex}$ = 3.761~MeV, ln(M) = 1.325. Figure~17(a) considers only the first of these two levels,
 at  E$_{ex}$ = 3.587~MeV which is indicated to have J = 1 or 2. It corresponds to the fifth mark for J = 1 as well as for J = 2. Figure~17(a) shows that the J = 1 value is preferred. 
 
 Then the two successive uncertain spin levels at 
E$_{ex}$ = 3.680~MeV and E$_{ex}$ = 3.761~MeV have respectively J = 1 or 2, and J = 2 or 3. Figure 17(b), obtained after removing the E$_{ex}$ = 3.587~MeV from the distribution of J = 2,  shows that the spin of the E$_{ex}$ = 3.680~MeV is uncertain. It may be either J =1 or 2, but if it is 1, then the spin  J = 2 is unlikely for the E$_{ex}$ = 3.761~MeV level.

In conclusion  for these three  unknown spin level, the most probable identification may be: J~=~1 for the E$_{ex}$ = 3.587~MeV and the E$_{ex}$ = 3.680~MeV levels, and J~=~3 for the E$_{ex}$ = 3.761~MeV level.
\subsection{The $^{24}_{12}$Mg nucleus}
\begin{figure}[ht]
\begin{center}
\hspace*{-3.mm}
\vspace*{5.mm}
\scalebox{1.05}[1]{
\includegraphics[bb=42 132 520 545,clip,scale=0.5]{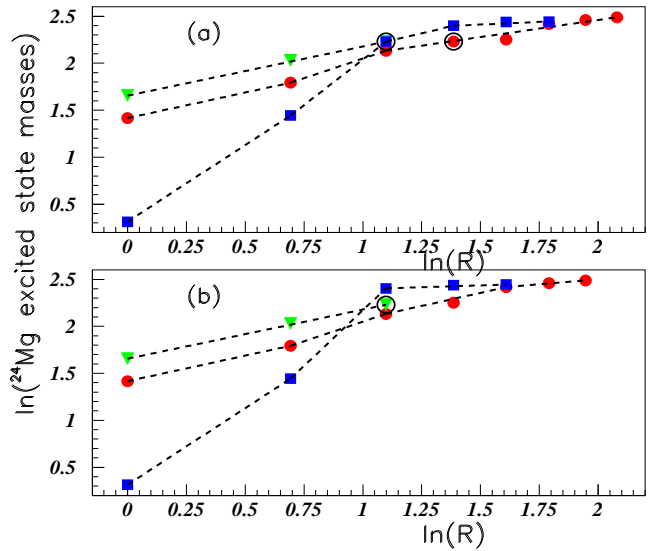}}
\caption{Color on line. Log-log plot of $^{24}$Mg excited level masses (see text). Full red circles correspond to J = 4 levels,  full blue squares correspond to J = 2 levels,
 and full green triangles correspond to J~=~3 levels.}
\end{center}
\end{figure}
Figure~18 shows the log-log plot of $^{24}$Mg  excited level masses \cite{134ba} versus the corresponding rank "R".  Except for the ground state level (J~=~0), the spins are J = 2, 3, or 4. In the mass range considered here, only two levels have J~=~3 and one level has J~=~1.  In figure~18, full red circles correspond to J~=~4 levels,  full blue squares correspond to J = 2 levels, and full green triangles correspond to J~=~3 levels. The spin of the level at E$_{ex}$=9.30~MeV, ln(M) = 2.230, is uncertain, since noted as (4)$^+$.  Figure~18(a)  shows the log-log distributions including the E$_{ex}$=9.30~MeV level (encircled mark) in the three possible spins. The linearity is worth for J~=~2. Figure~18(b) shows the log-log distribution with the assumption that the spin of the E$_{ex}$=9.30~MeV level is J~=~3. The linearity for J~=~3 remains very good, and is not modified for J~=~4. 

In conclusion the spin of  E$_{ex}$=9.30~MeV level is likely to be J~=~3.

\section{Discussion}
The fractal property, namely the straight line in the plot of log of successives masses versus the log of the corresponding rank,
is used for definite spin distributions:\\
 - to tentatively predict the spin of several mesons, and baryons, when this spin is unknown,\\
  - to tentatively predict the unknown spins of some nuclei excited level masses.
  
The different examples discussed above have shown that this method is often unable to give a firm result. Indeed, it needs the knowledge of the spin of several lower masses. 
It works also better when several masses larger than the studied mass, have known spins. But, above all, it can help to select the spin,  when the masses of the levels are not too close.

 The method is unappropriate if used for  rather large level ranks. Indeed, then, the masses are brought together leading to a large uncertainty. The method may be more suitable for light unstable nuclei. 
\section{Acknowledgments}
Ivan Brissaud introduces me to the study of the hadronic masses inside fractals. I thank him for his stimulating remarks and interest.

\pagestyle{plain}
\begin{table*}[t]
\begin{center}
\caption{Color on line. Meson masses (in MeV) and spins. Only first masses are considered. The masses of mesons with unknown spin are in red.}
\label{Table I}
\begin{tabular}[t]{c c c}
\hline
meson&J& masses (MeV)\\
\hline
Unflavoured&0&137.2, 475, 547.853, 957.78, 980, 980, 1294, 1300, 1350, 1409.8, 1474\\
                    &0&1476, 1505, 1720, 1756, 1816,  \textcolor{red}{1833.7}, 1992, 2103, 2189, 2226, 2330\\
                    &1& 775.49, 782.65, 1019.455, 1170, 1229.5, 1230, 1281.8, 1354, 1386, 1425, 1426.4\\
                    &1& 1465, 1518, 1570, 1594, 1647, 1662, 1670, 1680, 1720,  \textcolor{red}{1833.7}, 1900, 2149, 2175\\ 
                    &2&1275.1, 1318.3, 1430, 1525, 1562, 1617, 1639,, 1672.4, 1732, 1815\\
                    &2& \textcolor{red}{1833.7}, 1842, 1895, 1903, 1944, 2011, 2090, 2157,  \textcolor{red}{2231.1}, 2297, 2339\\
                    &3&1667, 1688.8, 1833.7, 1854, 1982, 2250\\
                    &4&1833.7, 2001, 2018,  \textcolor{red}{2231.1}, 2300\\
strange&0&495, 800, 1425, 1460, \textcolor{red}{1630}, 1830, 1945, \textcolor{red}{3100}\\
             &1&891.66, 1272, 1403, 1414, \textcolor{red}{1630}, 1650, \textcolor{red}{3100}\\
             &2&1425.6, 1580, \textcolor{red}{1630}, 1773, 1816, 1973, 2247, \textcolor{red}{3100}\\
             &3&\textcolor{red}{1630}, 1776, 2324, \textcolor{red}{3100}\\
             &4&\textcolor{red}{1630}, 2045, 2490, \textcolor{red}{3100}\\                     
charmed&0&1867.2, 2318, 2403, \textcolor{red}{2637}\\
                 &1&2008.6, 2422.7, 2427, \textcolor{red}{2637}\\
                 &2&2463.5, \textcolor{red}{2637}\\
char.-stra.&0&1968.47, \textcolor{red}{2112.3}, 2317.8, \textcolor{red}{2572.6}\\
                 &1&\textcolor{red}{2112.3}, 2459.5, 2535.29, \textcolor{red}{2572.6}, 2709\\
$c-{\bar c}$&0&2980.3, 3414.75, 3637, \textcolor{red}{3871.56}, \textcolor{red}{3915.5}\\
                    &1&3096.916, 3510.66, 3525.42, 3686.09, 3772.92, \textcolor{red}{3871.56}, \textcolor{red}{3915.5}\\
                    &2&3556.2, \textcolor{red}{3871.56}, \textcolor{red}{3915.5}, 3929\\                
\hline
\end{tabular}
\end{center}
\end{table*}

\pagestyle{plain}
\begin{table*}[t]
\begin{center}
\caption{Color on line. Baryon masses (in MeV) and spins.  Only first masses are considered. The masses of baryons with unknown spin are in red.}
\label{Table II}
\begin{tabular}[t]{c c c}
\hline
baryon&J&masses (MeV)\\
    N      &1/2&939, 1440, 1535, 1655, 1710, 2090, 2100\\
             &3/2&1520, 1700, 1720, 1900, 2080\\
             &5/2&1675, 1685, 2000, 2200\\
             &7/2&1990, 2190\\
             &9/2&2250, 2275\\
$\Delta$&1/2&1630, 1750, 1900, 1910, 2150\\
              &3/2&1232, 1600, 1700, 1920, 1940\\
              &5/2&1890, 1960, 2000, 2350\\
              &7/2&1930, 2200, 2390\\
              &9/2&2300, 2400\\
$\Lambda$&1/2&1115.683, 1406.5, 1600, 1670, 1800, 1810, \textcolor{red}{2000}, \textcolor{red}{2585}\\
               &3/2&1519.5, 1690, 1890, \textcolor{red}{2000}, 2325, \textcolor{red}{2585}\\
               &5/2&1820, 1830, \textcolor{red}{2000}, 2110, \textcolor{red}{2585}\\
               &7/2&\textcolor{red}{2000}, 2020, 2100, \textcolor{red}{2585}\\
                &9/2&\textcolor{red}{2000}, 2350, \textcolor{red}{2585}\\  
$\Sigma$&1/2&1193, \textcolor{red}{1480}, \textcolor{red}{1560}, 1620, \textcolor{red}{1620}, 1660, \textcolor{red}{1670}, \textcolor{red}{1690}, 1750, 1770\\
                &1/2&1880, 2000, \textcolor{red}{2250}, \textcolor{red}{2455}, \textcolor{red}{2620}, \textcolor{red}{3000}, \textcolor{red}{3170}\\
                &3/2&1382.8, \textcolor{red}{1480}, \textcolor{red}{1560}, 1580, \textcolor{red}{1620}, 1670, \textcolor{red}{1670}, \textcolor{red}{1690}, 1840\\
                &3/2&1940, 2080, \textcolor{red}{2250}, \textcolor{red}{2455}, \textcolor{red}{2620}, \textcolor{red}{3000}, \textcolor{red}{3170}\\
                &5/2&\textcolor{red}{1480}, \textcolor{red}{1560}, \textcolor{red}{1620}, \textcolor{red}{1670}, \textcolor{red}{1690}, 1775, 1915, 2070, \textcolor{red}{2250}, \textcolor{red}{2455}\\
                &5/2&\textcolor{red}{2620}, \textcolor{red}{3000}, \textcolor{red}{3170}\\
                &7/2&\textcolor{red}{1480}, \textcolor{red}{1560}, \textcolor{red}{1620}, \textcolor{red}{1670}, \textcolor{red}{1690}, 2030, 2100, \textcolor{red}{2250}, \textcolor{red}{2455}, \textcolor{red}{2620}, \textcolor{red}{3000}, \textcolor{red}{3170}\\                               
                &9/2&\textcolor{red}{1480}, \textcolor{red}{1560}, \textcolor{red}{1620}, \textcolor{red}{1670}, \textcolor{red}{1690}, \textcolor{red}{2250}, \textcolor{red}{2455}, \textcolor{red}{2620}, \textcolor{red}{3000}, \textcolor{red}{3170}\\
\hline
\end{tabular}
\end{center}
\end{table*}
\pagestyle{plain}
\begin{table*}[t]
\begin{center}
\caption{Color on line. Excitation energies (in MeV) and spins of discussed excited levels.  Only first masses are considered. The unknown spin levels are in red.}
\label{Table III}
\begin{tabular}[t]{c c c}
\hline
nucl.&J& masses (MeV)\\
\hline
$^4$He&0&0, 20.2, 21.4, \textcolor{red}{24.9}, \textcolor{red}{26.5}\\
              &1&\textcolor{red}{25.1}, \textcolor{red}{25.8}, 27.8, 28.3\\
              &2&22.4, 24.3, 28.5\\           
$^{9}$Be&1/2&1.684, 2.78, \textcolor{red}{13.79}, \textcolor{red}{15.97}\\      
                 &3/2&0, 4.704, 5.59, \textcolor{red}{13.79}, \textcolor{red}{15.97}\\                    
                 &5/2&2.429, 3.049, 7.94, 11.81, \textcolor{red}{13.79}, \textcolor{red}{15.97}, 16.671\\
                 &7/2&6.38, 11.283, \textcolor{red}{13.79}, \textcolor{red}{15.97}, 17.493\\ 
$^{13}$C&1/2&0, 3.0884, 8.86, 10.996, 11.080, \textcolor{red}{11.748}\\
                  &3/2&3.68437, 7.677, 8.2, 9.498, 9.897, \textcolor{red}{11.748}, 11.851\\
                  &5/2&3.85362, 6.864, 7.547, 10.818, \textcolor{red}{11.748}\\
                  &7/2&7.492, 10.753, \textcolor{red}{11.748}\\                                  
$^{14}$N&1&0, 3.9478, 5.6896, 6.2035, 8.062, 9.703, 10.228\\
              &2&5.1059, 7.0279, 7.9666, 8.979, 9.129, 9.1708, \textcolor{red}{9.386}, 9.509, 10.101, 10.434\\
              &3&5.8324, 6.444, 8.9091, 9.1241, \textcolor{red}{9.386}\\
$^{15}$N&1/2&0, 5.29887, 8.31279, 9.05, 9.225, \textcolor{red}{9.928}, 11.2929, 11.4375, 11.615, 11.965\\
                 &3/2&  6.32385, 7.30109, 8.5714, 9.15224, \textcolor{red}{9.928}, 10.07, 10.7019, 10.804, \textcolor{red}{11.235}, 11.778, 11.876, 12.145\\ 
                 &5/2&5.2704, 7.15536, 9.15527, 9.76, 10.4497, 10.5333, \textcolor{red}{11.235}, 12.095\\             
$^{16}$N&1&0.39727, 3.3528, 4.3204, 4.3914, 4.76, 5.318\\ 
              &2&0, 3.5227, 4.7828, 5.0537, \textcolor{red}{5.129}, \textcolor{red}{5.150}, 5.25\\
               &3&0.29822, 3.9627, \textcolor{red}{5.129}, \textcolor{red}{5.150}, 5.2301, 5.5216\\
$^{17}$F&1/2&0.49533, 3.104, 5.682, 6.037, \textcolor{red}{6.406}, 6.56\\
                 &3/2&4.64, 5.0, 5.488, 5.82, \textcolor{red}{6.406}, 6.774, 7.356\\
                 &5/2&0, 3.857, 6.697, 7.027\\             
$^{20}$F&1&0.98371, 1.056818, 3.17258, 3.48849, \textcolor{red}{3.58656}, \textcolor{red}{3.68013}, 3.96519, 4.08208\\
                 &2&0, 1.30934, 1.84397, 2.04405, \textcolor{red}{3.58656},  \textcolor{red}{3.68013}, \textcolor{red}{3.7611}\\
                &3&0.656, 1.9708, 2.19436, 2.8649, 2.96616, \textcolor{red}{3.7611}\\
                &4&0.82268, 2.968\\
$^{24}$Mg&2&1.36859, 4.2385, \textcolor{red}{9.300}, 11.017, 11.457, 11.521\\
                    &3&5.236, 7.616, \textcolor{red}{9.300}\\
                    &4&4.1228, 6.0103, 8.436, \textcolor{red}{9.300}, 9.515, 11.22, 11.693, 12.05 \\
\hline
\end{tabular}
\end{center}
\end{table*}
\end{document}